\documentclass[pra,twocolumn,showpacs]{revtex4}
\usepackage[english]{babel}
\usepackage{amsmath}
\usepackage{amssymb}
\usepackage{epsfig}

\begin{document}
\title{Three-dimensional stability of leapfrogging quantum vortex rings}
\author{Victor~P. Ruban}
\email{ruban@itp.ac.ru}
\affiliation{L.D. Landau Institute for Theoretical Physics RAS, 
142432 Chernogolovka, Moscow region, Russia} 

\date{\today}

\begin{abstract}
It is shown by numerical simulations within a regularized Biot-Savart law 
that dynamical systems of two or three leapfrogging coaxial quantum vortex
rings having a core width $\xi$ and initially placed near a torus of radii
$R_0$ and $r_0$, can be three-dimensionally (quasi-)stable in some regions 
of parameters $\Lambda=\ln(R_0/\xi)$ and $W=r_0/R_0$. At fixed $\Lambda$,
stable bands on $W$ are intervals between non-overlapping main parametric
resonances for different (integer) azimuthal wave numbers $m$.  
The stable intervals are most wide ($\Delta W\sim$ 0.01--0.05) between 
$m$-pairs $(1,2)$ and $(2,3)$ at $\Lambda\approx$ 4--12 thus corresponding 
to micro/mesoscopic sizes of vortex rings in the case of superfluid $^4$He.
With four and more rings, at least for $W>0.1$, resonances 
overlap for all $\Lambda$ and no stable domains exist.
\end{abstract}
\maketitle

\section{Introduction}

\begin{figure}
\begin{center}
\epsfig{file=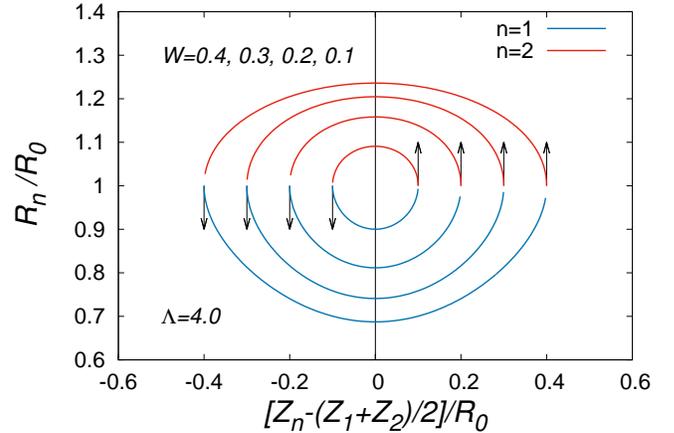, width=88mm}
\end{center}
\caption{Unperturbed leapfrogging of two coaxial vortex rings in the moving
frame of reference. Trajectories of vortices in an axial plane are closed, 
and they form cross sections of deformed tori. Both vortices move along the 
same trajectory, but shifted in time on a half-period. At given $W$, the 
trajectory passes points $z/R_0=\pm W$, $r/R_0=1$.}
\label{torus} 
\end{figure}

Dynamics of coaxial vortex rings in a fluid was considered by many authors
starting from Helmholtz (see, e. g., \cite{H1858,SL1992,M2010,BKM2013}, 
and numerous references therein). In ordinary fluid, a finite-core vortex 
ring is subjected to short-scale instabilities which deform cross section of 
the core and prevent the ring from a distant propagation 
(see \cite{WBT1974,W1975,WT1977,HF2003,FH2005}, and references therein). 
In contrast, quantized vortices in superfluid $^4$He at low temperature 
(helium-II) behave much like purely one-dimensional objects and are known to be 
intrinsically stable (because  quantum vortex rings and filaments in $^4$He 
do exist and are observed experimentally for macroscopic times
\cite{RR1964,YGP1979,BS2009,WTZG2014}). 
Therefore, instabilities can occur only when $N\geqslant 2$ quantum rings 
interact. In this context, a particularly interesting dynamical regime is 
the so called leapfrogging motion of two or more coaxial vortex rings when
they periodically pass through one another infinitely many times,
as exemplified in Fig.\ref{torus} (for a review, see \cite{M2010,BKM2013}). 
While investigated in detail for idealized axially symmetric configurations, 
interacting quantum vortex rings with distortions in three dimensions were 
theoretically considered just in a few recent works
\cite{CTCK2014,WBB2014,ZEBWG2016,GNP2015,GHP2016}. 
In particular, perturbations of vortex ring pairs were studied analytically
in \cite{GNP2015,GHP2016}, but with parameters corresponding to short-scale 
instabilities. Numerical simulations of macroscopic-size toroidal vortex
bundles in Ref. \cite{WBB2014} demonstrated development of three-dimensional
(3D) instabilities. In the last case the ratio of central torus radius 
$R_0$ to a vortex core width $\xi$ was of order $R_0/\xi\sim 10^8$, thus
corresponding to value of the local induction parameter 
$\Lambda=\ln(R_0/\xi)\approx 18$. 

In the present work, it will be demonstrated numerically that systems of two
or three leapfrogging coaxial quantum vortex rings with not so large 
$\Lambda=$ 4--12 can be 3D-stable and propagate almost unchanged over many
hundreds of initial diameters. Whether the motion is stable at given $\Lambda$, 
it strongly depends on another parameter, $W=r_0/R_0$, where $r_0$ is a poloidal 
(smaller) torus radius describing initial vortex configuration. A general 
explanation of the stability origin is the following. Let shapes of perturbed 
vortex rings be given in polar coordinates as $N$ pairs of functions
\begin{eqnarray}
Z_n(\varphi,t)&=&Z^{(0)}_n(t)+\mbox{Re}\sum_{m\geqslant 1}z_n^{(m)}(t)
\exp({\rm i}m\varphi),\\
S_n(\varphi,t)&=&S^{(0)}_n(t)+\mbox{Re}\sum_{m\geqslant 1}s_n^{(m)}(t)
\exp({\rm i}m\varphi),
\end{eqnarray}
where $n=1,\dots,N$, and $S_n(\varphi,t)=R^2_n(\varphi,t)/2$ is canonically
conjugate to $Z_n(\varphi,t)$. The crucial point is that 3D instabilities are
parametric in their nature, because linearised equations of motion for
$z_n^{(m)}(t)$ and $s_n^{(m)}(t)$  contain (real) time-dependent coefficients
depending on $Z^{(0)}_n(t)$ and $S^{(0)}_n(t)$,
\begin{eqnarray}
 \dot z_n^{(m)}&=&\sum_{n'}[A^{(m)}_{n,n'}(t)s_{n'}^{(m)} 
 + B^{(m)}_{n,n'}(t)z_{n'}^{(m)}],
 \label{z_t}\\
-\dot s_n^{(m)}&=&\sum_{n'}[B^{(m)}_{n',n}(t)s_{n'}^{(m)} 
+ C^{(m)}_{n,n'}(t)z_{n'}^{(m)}],
\label{s_t}
\end{eqnarray}
where  matrices $A^{(m)}$ and $C^{(m)}$ are symmetric, with dominating 
(at $m\geqslant 2$, $\Lambda\gg 1$)
diagonals, while $B^{(m)}$ has zeroes everywhere on the diagonal. Consider 
for the moment a purely periodic unperturbed leapfrogging with a time period
$T$. Then all the coefficients in  Eqs.(\ref{z_t})-(\ref{s_t}) are time-periodic
with this period. It is always the case for $N=2$, but for $N=3$ there also 
exist periodic solutions (so called relative choreographies \cite{BKM2013}). 
We will imply that the rings are on a (stationary moving) deformed torus having 
a $z$-size $2r_0$ and a mean $r$-size $R_0$ from $z$-axis (see Fig.\ref{torus}). 
Previous simulations have shown that period $T$ of relative axisymmetric motion 
depends on $\Lambda$ and $W$ \cite{M2010,BKM2013}. 
According to general mathematical theory, characteristic multipliers of our 
linear Hamiltonian system (\ref{z_t})-(\ref{s_t}) with time-periodic coefficients 
have the form
\begin{equation}
\rho_{\pm,i}^{(m)}=\exp\Big(\pm\sqrt{\mu_i^{(m)}}T\Big), 
\qquad i=1,\dots, N,
\end{equation}
where $\mu_i^{(m)}(\Lambda,W)$ are real functions. If at least one of them is
positive, then the motion is parametrically unstable at given $\Lambda$ and 
$W$. Qualitatively, this phenomenon is similar to well-known parametric 
instabilities for a single oscillator described by equation 
$\eta_{\tau\tau}+[\alpha+\delta\cos(\tau)]\eta=0$ with parameters $\alpha$ and 
$\delta$ (parametric resonances take place near $\alpha=(p/2)^2$, with $p=1,2,\dots$,
while the corresponding increments $\gamma_p\sim\delta^p$).
Let us in our case introduce a discrete index $\tilde p$ which enumerates positive
``humps''of $\mu_{\max}^{(m)}(\Lambda,W)$ at a fixed $\Lambda$. Among them there are
the most ``dangerous'' (main) parametric resonances. Our simulations for $N=2$
and $N=3$ (in the last case the unperturbed relative motion was actually only
approximately periodic) have shown that at relatively small 
$\Lambda\lesssim 3$, neighboring main parametric resonances overlap on
parameter  $W$. With larger $\Lambda=$ 4--8 however, intervals between
resonances appear at $W\approx$ 0.2--0.25 and reach widths of order
$\Delta W\sim$ 0.01--0.05 (the intervals are more prominent for $N=2$). 
At even larger  $\Lambda\gtrsim 15$ (macroscopic sizes of rings), stable 
intervals are shifted towards smaller $W\lesssim 0.15$ (thin tori).

Strictly speaking, secondary parametric resonances also contribute to overall
instability, especially for $N=3$, but they are weaker and not so wide to
close stable ``windows'' completely.

The structure of main instabilities in the case of coaxial rings turned out 
to be relatively simple, in contrast to recently observed numerically but yet
unexplained analytically, qualitatively similar instabilities of torus quantum
vortex knots and links \cite{R2018-1,R2018-2}.

\section{Approximate dynamical model}

\begin{figure}
\begin{center}
a)\epsfig{file=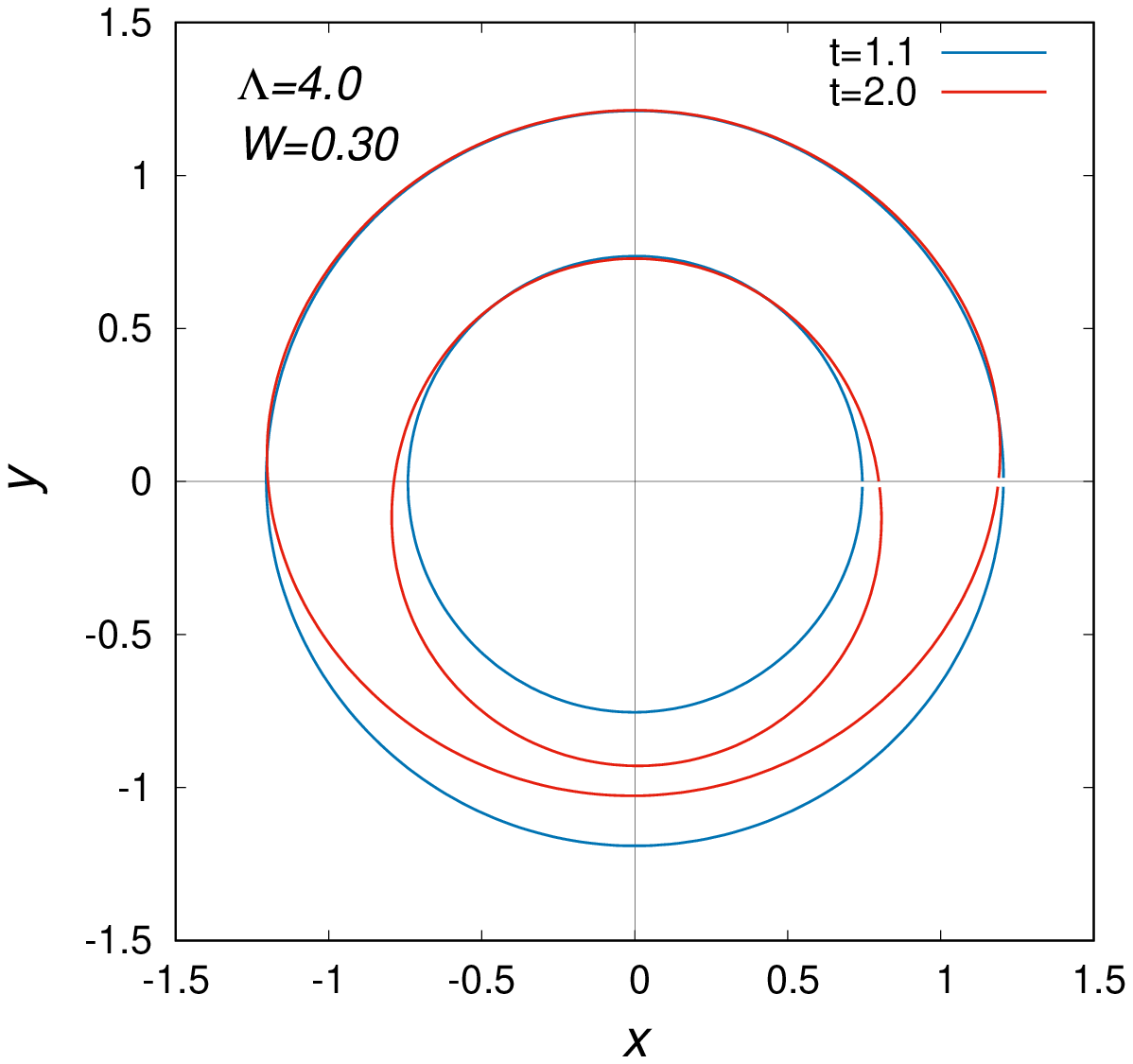, width=74mm}\\
b)\epsfig{file=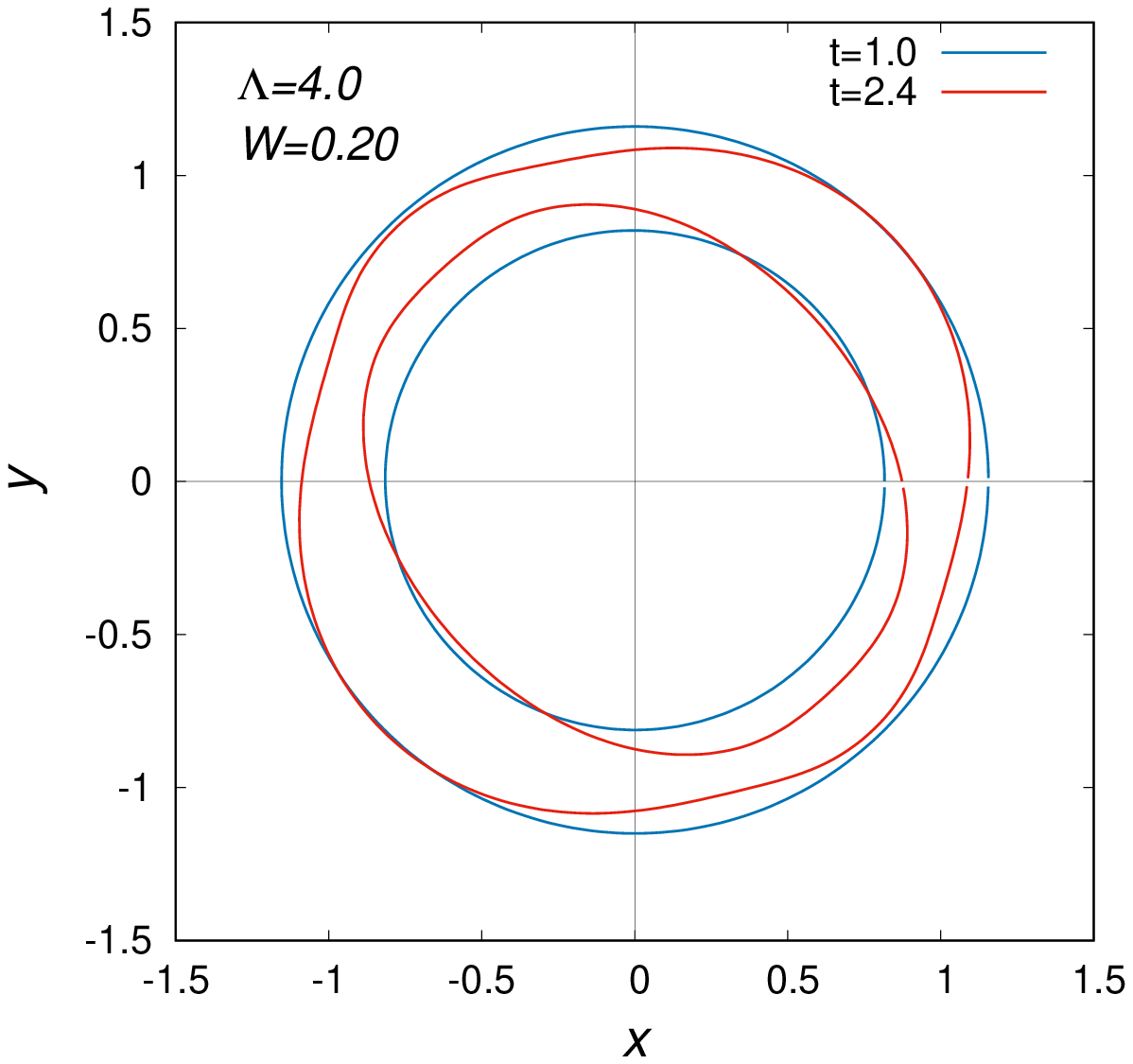, width=74mm}\\
c)\epsfig{file=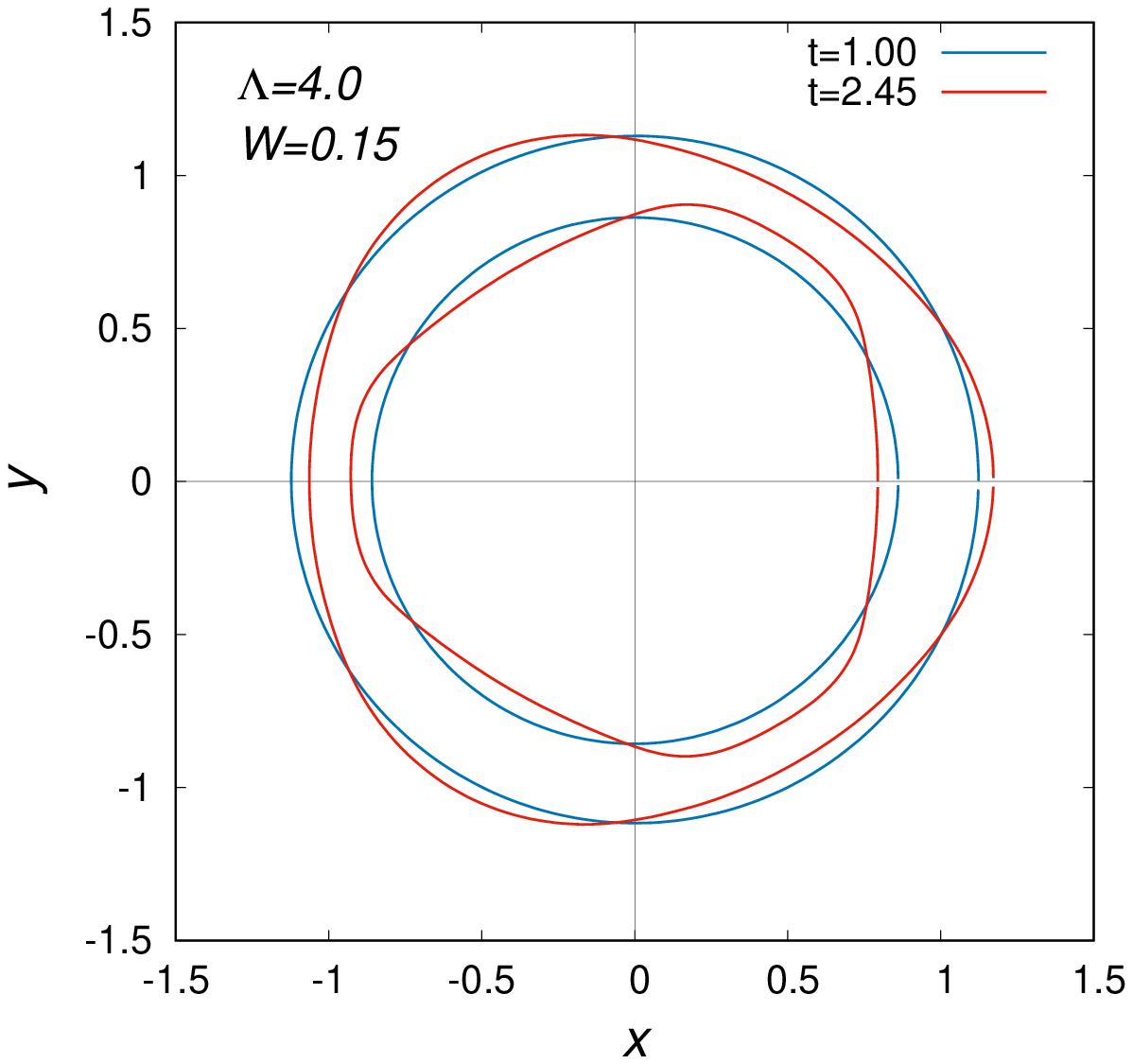, width=74mm}
\end{center}
\caption{Examples of instability development in a system of two vortex rings 
for: a) $m=1$; b) $m=2$; c) $m=3$.}
\label{m1m2m3} 
\end{figure}

Apparently, the main characteristics of instabilities are functions 
$\mu_i^{(m)}(\Lambda,W)$. 
In general, they could be extracted by a complicated procedure
from analysis of linearized equations of motion (\ref{z_t})-(\ref{s_t}). 
However, in this work we actually do not take that way, since matrices  
$A^{(m)}(t)$, $B^{(m)}(t)$, and $C^{(m)}(t)$ are very cumbersome
expressions involving special functions even in the simplest case $N=2$
\cite{GNP2015,GHP2016}. Instead, we employ a different approach and 
simulate the motion of vortex rings numerically until their significant deformation, 
thus including nonlinear stages of instabilities into consideration (see three 
examples in Fig.\ref{m1m2m3}). At the end of each run, we obtain an
estimate for lifetime of vortex system at given values of initial parameters
(and we believe the estimate is close to what could be observed in a real situation). 
After that we collect data and present the results as plots for inverse lifetime 
depending on $W$, for different $\Lambda$. Our numerical experiments are based
on a simplified but quite accurate mathematical model, discussed below.

To describe dynamics of vortex rings in three dimensions, we treat them 
in a usual manner as closed smooth space curves determined by $N$ vector functions 
${\bf X}_n(\beta,t)$ depending on an arbitrary longitudinal parameter $\beta$ 
and time $t$. We adopt a regularized Biot-Savart law in the form 
(see \cite{L1985,R2018-1,R2018-2} and compare to \cite{S1985,TAN2000,BB2011})
\begin{equation}
\dot{\bf X}_n(\beta,t)\!=\!\sum_{j=1}^{N}\!\frac{\Gamma}{4\pi}
\!\oint \!\frac{\tilde{\bf X}_j'\times({\bf X}_n-\tilde{\bf X}_j)}
{\sqrt{(|{\bf X}_n\!-\!\tilde{\bf X}_j|^2\!+\!a^2)^3}}d\tilde\beta
+\frac{\Gamma\lambda}{4\pi}\varkappa_n{\bf b}_n,
\label{BS_reg}
\end{equation}
where  $\Gamma=2\pi\hbar/m_{\rm at}$ is the circulation quantum (with atomic
mass $m_{\rm at}$), $\tilde{\bf X}_j={\bf X}_j(\tilde\beta,t)$, 
$\tilde{\bf X}_j'=\partial {\bf X}_j(\tilde\beta,t)/\partial\tilde\beta$, 
$\varkappa_n(\beta,t)$ is a local curvature of the filament 
(in the usual geometric sense), and ${\bf b}_n(\beta,t)$ is a local unit binormal
vector on the curve. Here $\lambda$ is a (positive) stiffness parameter
characterizing potential energy of quantum vortex core, and 
$a=\xi\exp(\lambda)$ is a geometric core radius. Note that the last term in 
Eq.(\ref{BS_reg}) corresponds formally to the local induction approximation (LIA),
while the sum of nonlocal integrals is a smoothly regularized Biot-Savart 
law for vorticity distribution in the form of several infinitely thin filaments,
each having one quantum of circulation. Thus, we define an effective
core width $\xi$ in such a manner that the total local induction parameter is 
$\Lambda=\lambda+\ln(R_0/a)=\ln(R_0/\xi)$. 

The presence of two parameters,
$a$ and $\lambda$, makes the model quite flexible to quantitatively describe
relatively long-scale dynamics of real vortices in a superfluid.

For simplicity, we work with non-dimensionalized quantities, so that formally
$\Gamma=2\pi$, $R_0=1$.

The model under consideration is a Hamiltonian system, with non-canonical
structure $[{\bf X}'_n\times\dot{\bf X}_n]=\delta{\cal H}/\delta{\bf X}_n$, 
and the Hamiltonian functional
\begin{equation}
{\cal H}=\frac{1}{4}\sum_{n,j}\oint\!\oint\!\frac{({\bf X}'_n\cdot 
\tilde{\bf X}'_j)d\beta d\tilde\beta}
{\sqrt{|{\bf X}_n-\tilde{\bf X}_j|^2+a^2}}+
\frac{\lambda}{2}\sum_{n}\oint |{\bf X}'_n|d\beta.
\label{H}
\end{equation}
In this expression, the double sum of non-local terms corresponds to kinetic 
energy of superfluid flow created by quantized vortices, while the ordinary sum 
of local integrals is potential energy of the cores. 

Canonically conjugate functions are different in different curvilinear
coordinate systems. For example, when parameterized by the azimuthal angle 
in polar coordinates, $Z_n(\varphi, t)$ and  $R^2_n(\varphi, t)/2$ are
canonical variables. Matrices  $A^{(m)}$, $B^{(m)}$, and $C^{(m)}$ 
as functions of $Z^{(0)}_n$ and  $S^{(0)}_n$ then follow from expansion of 
the correspondingly expressed Hamiltonian in azimuthal perturbations up to 
the second order. We do not write them here.

Although physical applicability of the above hydrodynamical model is
restricted to relatively long scales $\tilde l\gg a$, but in numerical simulations
short scales of order $a$ can be also important.

\begin{figure}
\begin{center}
\epsfig{file=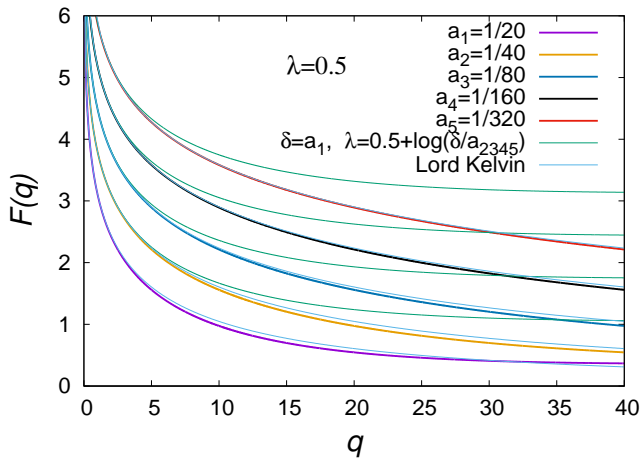, width=88mm}
\end{center}
\caption{Comparison between ``exact'' and approximate  dispersion laws 
for a straight vortex filament having a stiffness parameter $\lambda=0.5$.
Plots for $F(q)=2\omega_q/q^2$ are shown. Also curves corresponding to Lord 
Kelvin's dispersion relation (\ref{LK}) for $\nu=1$ and sign ``+''are drawn. 
The two expressions (\ref{lambda-a-model}) and  (\ref{LK}) are remarkably close 
to each other.}
\label{F_comparison} 
\end{figure}

It is parameter $\lambda$ that prevents short-scale instabilities (compare
with \cite{FS2001} where it was not included). Indeed, let us consider a
straight vortex filament. Small perturbations $X(z,t)$ and $Y(z,t)$ are
canonically conjugate in that case and described by quadratic Hamiltonian
\begin{eqnarray}
{\cal H}^{(2)}&=&\frac{1}{4}\int\int\frac
{( X'\tilde X'+ Y'\tilde Y')dz d\tilde z}
{\sqrt{(z-\tilde z)^2+a^2}}\nonumber\\
&-&\frac{1}{8}\int\int\frac
{[(X-\tilde X)^2+(Y-\tilde Y)^2]dz d\tilde z}
{\sqrt{[(z-\tilde z)^2+a^2]^3}}\nonumber\\
&+&\frac{\lambda}{4}\int (X'^2+Y'^2)dz.
\label{H_2}
\end{eqnarray}
The corresponding dispersion law is 
\begin{eqnarray}
\omega_q&=&\frac{q^2}{2}\left[\lambda + 
2K_0(qa)-\frac{2}{(qa)^2}+\frac{2K_1(qa)}{qa}\right]\nonumber\\
&=&\frac{q^2}{2}[\lambda+\tilde F(qa)],
\label{lambda-a-model}
\end{eqnarray}
with $K_0$ and $K_1$ being the modified Bessel functions. It should be noted
that different core models result in different expressions for $\omega_q$
(see, e.g., \cite{LK1880,MS1974,BHT2006}, and references therein). However,
with appropriate choice for $\lambda$, the present model is able to 
approximate them very closely. In particular, Lord Kelvin's \cite{LK1880}
dispersion relation for $\nu$-th azimuthal mode 
of a hollow-core columnar vortex can be represented as follows:
\begin{equation}
\omega^{L.K.}_{\nu,q}=\frac{1}{a^2}\Big[
\pm\Big(\nu+qa \frac{K_{\nu-1}(qa)}{K_\nu(qa)}\Big)^{1/2}-\nu\Big],
\label{LK}
\end{equation}
with $\nu=0,1,\dots$.
We are interested in the soft bending mode which corresponds to $\nu=1$
and sign ``+''. The value $\lambda=0.5$ works in this case very well 
(see Fig.\ref{F_comparison}). A more complicated dispersion equation for 
a constant-vorticity core \cite{LK1880} is not discussed here.

It is important that at small $u\ll 1$, function $\tilde F(u)\approx\ln (1/u)>0$, 
while at $u_0\approx 1.1$ it changes the sign, and at $u_*\approx 2.32$ 
it has a minimum where $\tilde F_{min}\approx -0.137$. Therefore, with zero 
or small $\lambda$, frequency $\omega_q=q^2[\lambda+\tilde F(qa)]/2$ 
is small somewhere at $qa\sim 1$.
When in a weak externally imposed transverse strain field [so that 
$\Delta {\cal H}^{(2)}=(\sigma/2)\int(X^2-Y^2)dz$], 
rectilinear vortex modifies its dispersion law, 
\begin{equation}
\Omega_q=\sqrt{\omega^2_q-\sigma^2},
\end{equation}
and instability can occur if the expression under square root becomes
negative (concerning a vortex filament in ordinary fluid under external
strain, see \cite{MS1975,TW1976,RS1984,F2003}).
In context of present work, a local strain on a ring is inevitably caused 
by its circular shape and by nonuniform velocity field from other rings.
Thus, a sufficiently large value of $\lambda>0.137$ is required to remove 
such instability.

We do not know a best fit for $\lambda$, say, in real superfluid $^4$He, 
but fortunately, it is not so important for our purposes. Indeed we need 
to simulate just relatively long-scale dynamics corresponding to 
$q\lesssim 10$ (because $q\sim m$, and we are interested in moderate $m$). 
In that domain, the dynamics depends mostly on the total
local induction parameter $\Lambda$, and does not ``feel'' actual value of 
$a$. Therefore one can replace parameters in the equations of motion
(\ref{BS_reg}) without significant loss of accuracy: 
\begin{equation}
a\to\delta, \qquad \lambda\to\lambda+\ln(\delta/a),
\label{subst}
\end{equation}
where $\delta>a$ (of course, vortex configuration should be far from mutual
and self-intersections). Fig.\ref{F_comparison} indicates that replacement
(\ref{subst}) only weakly changes dispersion law at moderate wave numbers. 
A direct way to see weak sensitivity of the system under (\ref{subst}) 
is to write in Eq.(\ref{H}):
\begin{eqnarray}
&&\frac{1}{\sqrt{|\Delta{\bf X}|^2\!+\!a^2}}=
\frac{1}{\sqrt{|\Delta{\bf X}|^2\!+\!\delta^2}}\nonumber\\
&&+\Big[\frac{1}{\sqrt{|\Delta{\bf X}|^2\!+\!a^2}}
-\frac{1}{\sqrt{|\Delta{\bf X}|^2\!+\!\delta^2}}\Big].
\end{eqnarray}
Since the expression in square bracket is effectively local, it approximately
results in $(1/2)\ln(\delta/a) \sum_n\oint |{\bf X}'_n|d\beta$, thus
confirming (\ref{subst}). In most simulations we used $\delta=1/20$, and only
for verification we used $\delta=1/40$ in some cases. Thus, the major
parameters of the system are $\Lambda$ and $W$, while $a$ 
plays a secondary role.

By the way, it is now clear why our model with $\lambda=1/2$ approximates 
the Lord Kelvin's formula for a hollow vortex in ordinary fluid. Indeed, 
since the core cross section tends to be uniform along the vortex at every
time moment, while the volume of core cavity is conserved, we have relation 
$b^2(t){\cal L}(t)=a^2{\cal L}_0$ between actual time-dependent core width $b(t)$ 
and vortex length ${\cal L}(t)=\int |{\bf X}'|d\beta$, with constants $a$ and 
${\cal L}_0$. Therefore the Hamiltonian of a single narrow hollow vortex is
\begin{equation}
{\cal H}_b\approx {\cal H}_a+\frac{1}{2}\ln\Big(\frac{a}{b}\Big){\cal L}
={\cal H}_a+\frac{1}{4}\ln\Big(\frac{\cal L}{{\cal L}_0}\Big){\cal L},
\end{equation}
where ${\cal H}_a$ is the $a$-regularized Biot-Savart Hamiltonian (the first 
term in Eq.(\ref{H})). Quadratic on $X$ and $Y$ part of the above  expression 
is exactly Eq.(\ref{H_2}) with $\lambda=1/2$.

\begin{figure}
\begin{center}
\epsfig{file=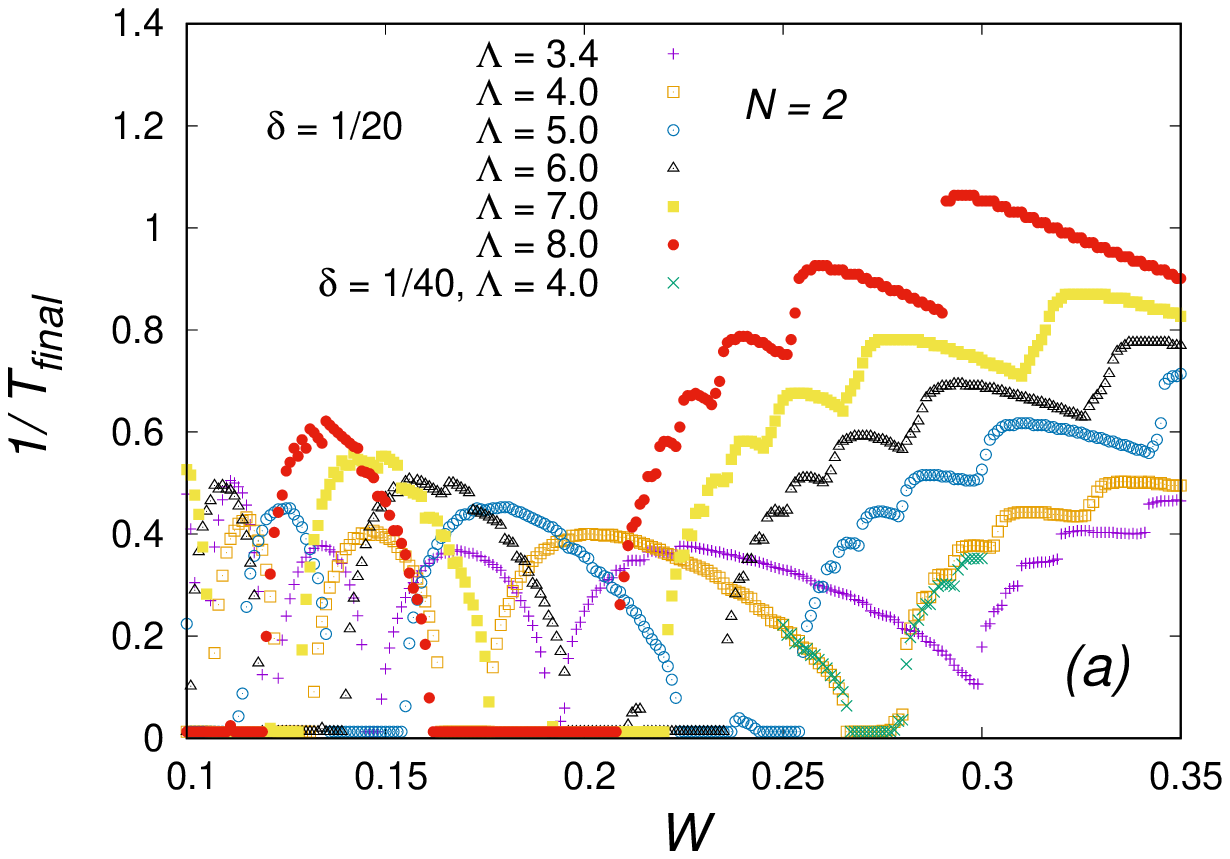, width=88mm}\\
\epsfig{file=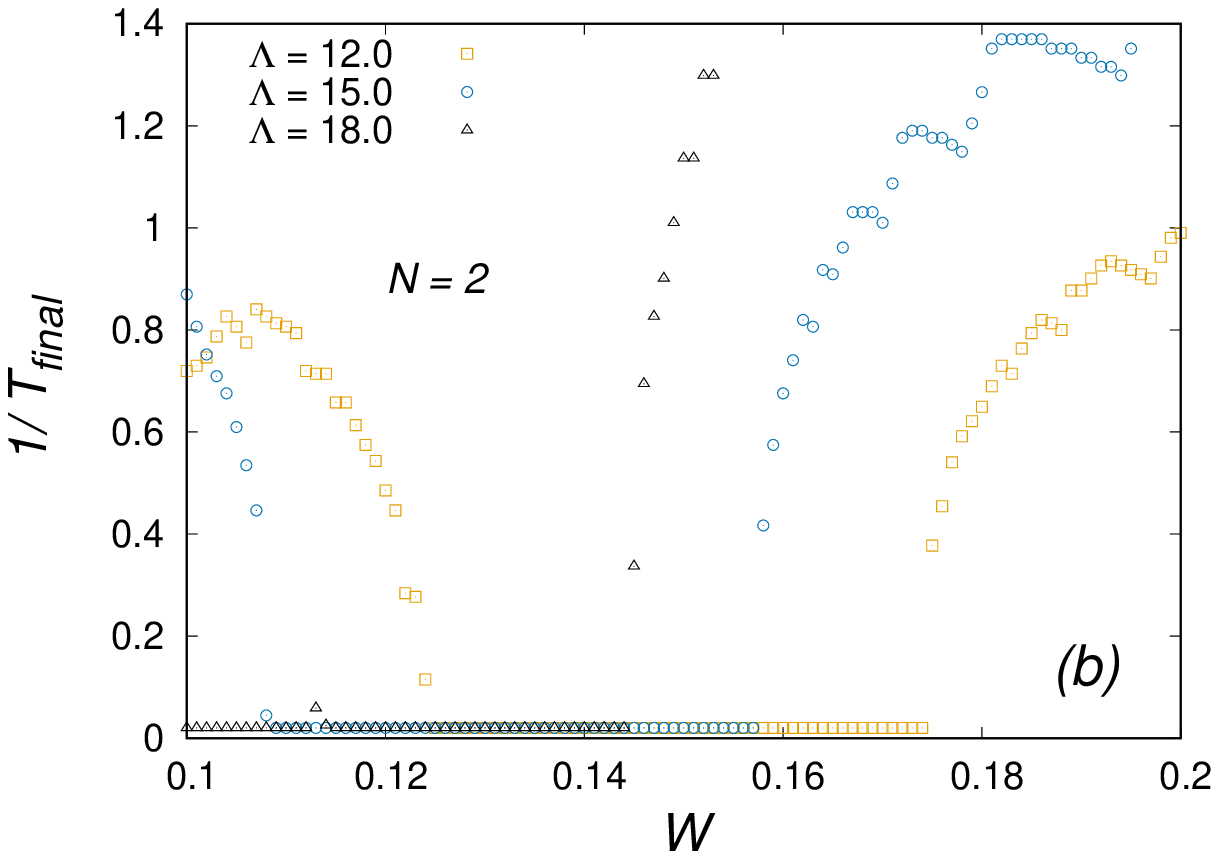, width=88mm}
\end{center}
\caption{Inverse lifetime for $N=2$ vortex rings: a) micro/mesoscopic sizes; 
b) almost macroscopic sizes (here $T_{\rm max}=50$). 
Relatively wide stable bands are seen. For a fixed $\Lambda$, 
the rightest big hump on the plot corresponds to main instability with 
$m=1$, the next one to the left is for $m=2$, and so on. 
Secondary instabilities are very weak.}
\label{N2} 
\end{figure}
\begin{figure}
\begin{center}
\epsfig{file=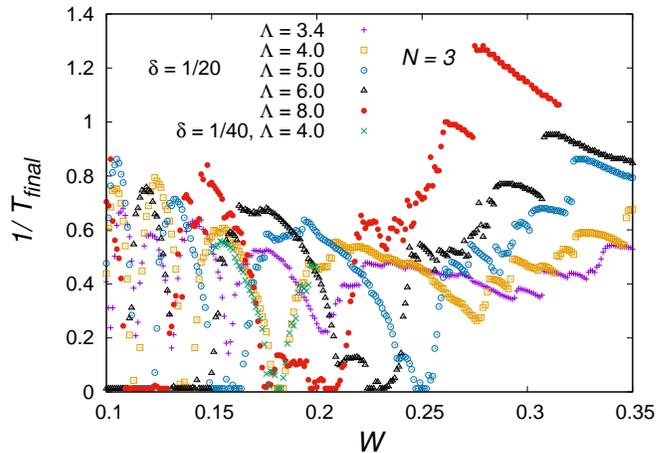, width=88mm}
\end{center}
\caption{Inverse lifetime for $N=3$ vortex rings. Here stable bands are not
so wide as for $N=2$, while secondary instabilities are stronger.}
\label{N3} 
\end{figure}

\section{Numerical results}

In numerical simulations, terms proportional to ${\bf X}_n'$ were added to 
equations of motion (\ref{BS_reg}), such that azimuthal component of $\dot{\bf X}_n$ 
was equal to zero. Such modification is allowed due to the freedom 
in choice of the longitudinal parameter $\beta$, and it controlled density 
of distribution of discrete points along the curve. 

The same procedure was used as described in 
Refs.\cite{R2018-1,R2018-2}. Basically, it was a pseudo-spectral scheme on 
the longitudinal parameter $\beta$,  in combination with a Runge-Kutta 4-th
order scheme for time stepping. Shape of each vortex filament was approximated 
by $L$ points ${\bf X}_{n,l}(t)={\bf X}_n(2\pi l/L,t)$ 
[typical values $L=256$ and $L=512$]. At that,
\begin{equation}
{\bf X}_{n,l}=\mbox{Re}\sum_{k=0}^{K-1}\hat{\bf X}_{n,k}\exp(2\pi {\rm i} kl/L).
\end{equation}
In each time step, $K\approx(3/8)L$ of the corresponding Fourier coefficients 
$\hat{\bf X}_{n,k}$ participated in the  Runge-Kutta procedure. After that, only
coefficients with $k$ not higher than $K_{\rm eff}\approx L/4$ were retained, 
while the remaining were set to zero. Such strategy has successfully proven 
itself in many different problems. In our case it also has shown good accuracy 
and stability.

For every run, an initial state consisted of
$N$ slightly perturbed, nearly coaxial vortex rings placed on a torus with
equal poloidal angles between them. More precisely, for $N=2$ it was a pair 
of equal rings with unit radius and a distance $2W$ between them. For $N=3$, 
it was $Z_n^{(0)}(0)=W\sin(2\pi n/3)$, $R_n^{(0)}(0)=1+W\cos(2\pi n/3)$
(of course, these values were not exactly corresponding to a choreography, 
but the resulting relative motion was nearly periodic, with small superimposed
additional oscillations). Typical amplitudes of initial 3D distortions were
0.001--0.01. The run was terminated either when the time reached $T_{\max}=80$, 
or when configuration of vortices became close to intersection. A final time
$T_{\rm final}$ was recorded, and it was taken as an estimate for vortex 
system lifetime. 

The data collected for $N=2$ and $N=3$ are presented in Fig.\ref{N2} and 
Fig.\ref{N3} respectively, which are essentially the main results of present
work. Obviously, stable domains, if any, should correspond to very small
values of the inverse lifetime. Indeed such intervals are clearly seen in 
the figures. It should be said that Hamiltonian was preserved typically up 
to 5-6 decimal digits in stable domains. In several additional simulations,
$T_{\max}$ was equal to 320, and in many cases no tendency to instability 
development was detected. Leapfrogging vortex rings thus traveled distances 
as many hundreds of initial diameters.

Especially clean are plots for $N=2$, because unperturbed relative motion 
was really periodic. What is interesting, in this case stability of 
a microscopic-size pair of rings can take place at rather large values
$W\approx 0.27$ (Fig.\ref{N2}a), when the corresponding ``torus'' is ``fat'' 
and strongly deformed. 
The widest stable intervals are between $m$-pairs (1,2) and (2,3), though they 
are opened not simultaneously as $\Lambda$ increases. For almost macroscopic 
rings with $\Lambda\approx$ 12--18, stable intervals are near $W\approx 0.15$ 
(Fig.\ref{N2}b).

For $N=3$, stable bands still exist, though not so wide and spoiled slightly
by secondary instabilities. Results for $N=4$ are not presented 
because there we did not observe stability zones.

\section{Conclusions}

Thus, accurate numerical simulations based on a widely accepted theoretical
model for nonlinear dynamics of quantum vortex rings have been carried out 
and demonstrated existence of parametrically stable regimes in leapfrogging
motion of two and three coaxial rings. The local induction parameter in
``optimal'' domain $\Lambda\approx$ 4--10 corresponds to micro/mesoscopic
sizes of vortex rings, if superfluid $^4$He is meant. In that domain, several
stable bands were found corresponding to rather appreciable relative distances 
between the rings (up to $2W\approx 0.55$ for the first stable interval). 

Further theoretical efforts should perhaps be directed towards immediate 
calculation of the monodromy matrix for linear system (\ref{z_t})-(\ref{s_t}) 
and finding its eigenvalues $\rho_{\pm,i}^{(m)}$.

A system of three vortex rings can exhibit also a more complicated regime 
as chaotic leapfrogging \cite{BKM2013}. That interesting case has not yet
been investigated for 3D stability.

The numerical results obtained here seem interesting and also potentially
stimulating for future experiments. Indeed, if one is going to investigate
parametrically stable nontrivial vortex configurations in a real-world
experiment, then the problem of controllable creation of a few mesoscopic-size
coaxial vortex rings can become in practice more easy than 
creation of vortex knots or links.

\begin{figure}
\begin{center}
\epsfig{file=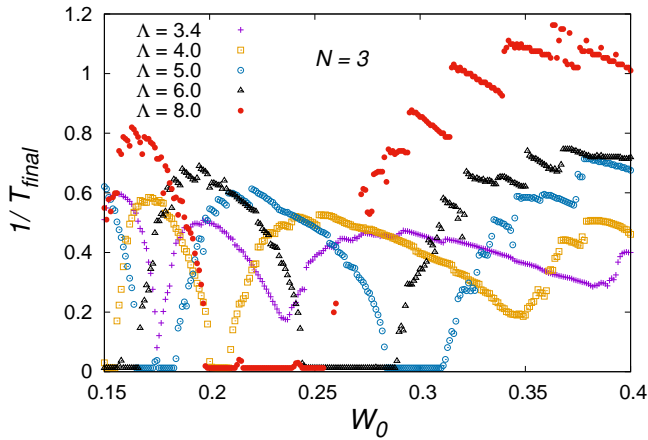, width=88mm}
\end{center}
\caption{\bf Inverse lifetime for $N=3$ vortex rings with initial positions 
closely corresponding to choreographies. Qualitatively, the plots are similar 
to Fig.\ref{N3}, but here stable intervals are quite wide.}
\label{N3_chor} 
\end{figure}

{\bf After acceptance of the above text for publication in Physics of Fluids,
additional simulations for three vortex rings were carried out with somewhat 
different,  empirically selected  initial ring positions much better 
corresponding to relative choreographies:
$Z_n^{(0)}(0)=W_0\sin(2\pi n/3)$, 
$R_n^{(0)}(0)=\sqrt{1+2W_0[1-D(\Lambda)W_0]\cos(2\pi n/3)}$.
Here coefficient $D(\Lambda)$ ranges from $\approx 1.0$ at $\Lambda=4.0$ 
to $\approx 1.16$ at $\Lambda=8.0$. Simulations have shown axial size $W$ 
of the resulting deformed torus to differ slightly from the auxiliary parameter 
$W_0$, so that $W=W_0-{\cal O}(W_0^2)$. What is important in this case,
it is the presence of more wide and clean stability intervals, presented 
in Fig.\ref{N3_chor}.
}

\end{document}